\documentclass{aa}

\usepackage{graphicx}
\usepackage{txfonts}
\def\lsim{\mathrel{\rlap{\lower4pt\hbox{\hskip1pt$\sim$}}
    \raise1pt\hbox{$<$}}}
\def\gsim{\mathrel{\rlap{\lower4pt\hbox{\hskip1pt$\sim$}}
    \raise1pt\hbox{$>$}}}
\begin{document}

   \title{Observational constraints on interacting dark matter model without dark energy}
   \author{Shuo Cao, Zong-Hong Zhu  \and Nan Liang  }
   \institute{Department of Astronomy, Beijing Normal   University, Beijing 100875, China\\
   \email{zhuzh@bnu.edu.cn,liangn@bnu.edu.cn}}

\abstract {} {The interacting dark matter (IDM) scenario allows for
the acceleration of the Universe without Dark Energy.}{We constrain
the IDM model by using the newly revised observational data
including $H(z)$ data and Union2 SNe Ia via the Markov chain Monte
Carlo method.}{ When mimicking the $\Lambda$CDM model, we obtain a
more stringent upper limit to the effective annihilation term at
$\kappa C_1\approx 10^{-3.4}\rm{Gyr}^{-1}$, and a tighter lower
limit to the relevant mass of Dark Matter particles at $M_x\approx
10^{-8.6}\rm{Gev}$. When mimicking the $w$CDM model, we find that
the effective equation of state of IDM is consistent with the
concordance $\Lambda$CDM model and appears to be most consistent
with the effective phantom model with a constant EoS for which
$w<-1$.} {}

\keywords{Cosmology: dark matter --- cosmological observations}

\authorrunning{Shuo Cao, Zong-Hong Zhu  and Nan Liang}
\titlerunning{Constraints on the IDM model}

\maketitle

%% The author head (on even pages) and the title head (on odd pages) will be
%% automatically extracted from \author{} and \title{}. Whenever the title is too long,
%% you will be asked to supply a shorter one by inserting either \authorrunning{} or
%% \titlerunning{} before \maketitle. Anyway, you can specify your own heads.
%%
%%
%% Note: In the following text body of your manuscript, please note several differences from
%%       other major journals:
%% (1) \subsection{Please Capitalize the First Letter of Each Notional Word in Subsection Title}
%% (2) Please Capitalize the First Letter of Each Notional Word in all tables' captions

%
%________________________________________________ sections below

%
%________________________________________________________________

\section{Introduction} \label{sec1}
Recent observations of type Ia supernovae (SNe Ia, Riess et al.
1998; Perlmutter et al.1999) have predicted that our present
universe is passing through an accelerated phase of expansion
proceeded by a period of deceleration. A new type of matter with
negative pressure, which is popularly known as dark energy, has been
proposed to explain the present phase of acceleration. The most
simple dark energy candidate, the cosmological constant
($\Lambda$CDM model), though known to be consistent with various
observations such as SNe Ia, the galaxy cluster gas mass fraction
data (Wilson et al. 2006; Davis et al. 2007; Allen et al. 2008), and
the CMB temperature and polarization anisotropies (Jassal et al.
2010), is always affected by the coincidence problem. Until now,
many other dark energy models have been brought forward to explain
this comic acceleration such as the scalar fields with a dynamical
equation of state [e.g., quintessence (Peebles \& Ratra 1988a,
1988b; Caldwell, Dave, \& Steinhardt 1998), phantom (Caldwell 2002),
k-essence (Armendariz-Picon et al. 2001), quintom (Feng et al. 2005;
Guo et al. 2005;  Liang et al. 2009)], the Chaplygin gas
(Kamenshchik et al. 2001; Bento et al. 2002), holographic dark
energy (Cohen 1999; Li 2004), and so on. Many alternatives to dark
energy in which gravity is modified have been proposed as a possible
explanation of the acceleration [e.g., the braneworld models (Dvali,
Gabadadze, \& Porrati 2000; Zhu \& Alcaniz 2005), the Cardassian
expansion model (Freese \& Lewis 2002, Zhu \& Fujimoto 2002)], as
well as the $f(R)$ theory (Capozziello \& Fang 2002; Carroll et al.
2004) and the $f(T)$ theory (Bengochea \& Ferraro 2009; Wu \& Yu
2010).

It has been shown that the dark matter self-interactions could
provide the accelerated expansion of the Universe without any dark
energy component (Zimdahl et al. 2001; Balakin et al. 2003). In the
framework of the Boltzmann formalism, if there is a disequilibrium
between the dark matter particle creation and annihilation
processes, an effective source term with negative pressure could be
created. Basilakos \& Plionis (2009) investigated the circumstances
under which the analytical solution space within the framework of
the interacting dark matter (IDM) scenario allows for a late
accelerated phase of the Universe, and find that the effective
annihilation term  of the simplest IDM model is quite small by using
the nine observational $H(z)$ data points of Simon et al. (2005).
The gravitational matter creation model that is fully dominated by
cold dark matter (Lima et al. 2008; Basilakos \& Lima  2010) is
mathematically equivalent to one case of the IDM models (Basilakos
\& Plionis 2009).

In this paper, we use the newly revised $H(z)$ data (Stern et al.
2010; Gazta$\tilde{n}$aga et al. 2009) and the Union2 set of 557 SNe
Ia (Amanullah  et al. 2010) to constrain the relevant IDM models
(Basilakos \& Plionis 2009) by using the Markov chain Monte Carlo
(MCMC) method. This paper is organized as follows. In Sect.
\ref{sec2}, we briefly indicate the basic equations of the IDM
models. Observational data including $H(z)$ and SNe Ia are given in
Sect. \ref{sec3}. In Sect. \ref{sec4}, MCMC constraint results from
different combined data sets are illustrated. Finally, we summarize
our main conclusions in Sect. \ref{sec5}.

\section{The basic equations of
the IDM models} \label{sec2} We assume that the IDM density obeys
the collisional Boltzmann equation (Kolb \& Turner 1990)
\begin{equation}
\label{1}
  \dot{\rho}+3H\rho+\kappa \rho^{2}-\Psi=0 \;,
\end{equation}
where $\Psi$ is the rate of creation of DM particle pairs and the
annihilation parameter $\kappa = \langle \sigma u \rangle /M_x$
(where $M_x$ is the mass of the DM particle, $\sigma$ is the
cross-section for annihilation, and $u$ is the mean particle
velocity). Compared to the usual fluid equation, the effective
pressure term is
\begin{equation}\label{3}
p_{\rm{eff}}=(\kappa \rho^2-\Psi)/3 H\;.
\end{equation}
When the IDM particle creation term is larger than the annihilation
term ($\kappa \rho^{2}-\Psi<0$), IDM may serve as a negative
pressure source in the global dynamics of the Universe (Zimdahl et
al. 2001; Balakin et al. 2003). Basilakos \& Plionis (2009)
phenomenologically identified two functional forms for which the
previous Boltzmann equation can be solved analytically, only one of
which is of interest since it indicates the dependence of the scale
factor on a ``$\propto a^{-3}$''. We refer to appendix B in
Basilakos \& Plionis (2009) for details. We assume that
\begin{equation}
\Psi(a)=a H(a)R(a)=C_{1}(n+3)a^{n} H(a)+\kappa C_{1}^{2}a^{2n} \;.
      \label{7}
\end{equation}
And the total energy density is
\begin{equation}
\label{8} \rho(a)=C_{1}a^{n}+a^{-3}\frac{F(a)}
{\left[C_2-\int_{1}^{a} x^{-3} f(x) F(x)dx\right]} \;,
\end{equation}
where $n$, $C_{1}$, and $C_2$ are the corresponding constants of the
problem ($\kappa C_1$ in the unit of Gyr$^{-1}$), and the kernel
function $ F(a)={\rm exp} [-2 \kappa C_{1}\int_{1}^{a}
x^{n-1}/{H(x)}dx] $. The first term on the right side of
Eq.(\ref{8}) obviously corresponds to the residual matter creation
that results from the possible disequilibrium between the IDM
particle creation and annihilation processes, while the second term
can be viewed as the energy density of the self IDM particles that
are dominated by the annihilation process.

 \subsection{ Model 1:  Mimicking the $\Lambda$CDM Model}
If $n = 0$, the global density evolution can be transformed as
\begin{equation}
\label{13} \rho(a)=C_{1}+a^{-3}\frac{e^{-2 \kappa C_{1} (t-t_0)}}
{\left[C_{2}- \kappa Z(t) \right]} \;\;,
\end{equation}
where $Z(t)=\int_{t_{0}}^{t} a^{-3}e^{-2 \kappa C_1 (t-t_0)}$
(Basilakos \& Plionis 2009). At the present epoch, the density
evolves according to $\rho(a)\simeq C_{1}+{a^{-3}}/{C_{2}},$ which
is approximately equivalent to the corresponding evolution in the
$\Lambda$CDM model, in which the $C_{1}$ term resembles the
cosmological constant term ($\rho_{\Lambda}$) and the $1/C_{2}$ term
resembles the density of matter ($\rho_m$). The Hubble parameter can
be written as (Basilakos \& Plionis 2009)
 \begin{equation}\label{15}
\left(\frac{H}{H_0}\right)^2 = \Omega_{1,0} + \Omega_{2,0} a^{-3}
\frac{ e^{-2 \kappa C_{1} (t-t_0)} }{1 + \kappa C_{1}
(\Omega_{2,0}/\Omega_{1,0}) Z(t)} \;,
\end{equation}
where $\Omega_{1,0}= 8 \pi G C_{1}/3H_0^2$ and $\Omega_{2,0}=8 \pi G
/ 3H^2_0 C_{2}$, which relate to $\Omega_{\Lambda}$ and $\Omega_{\rm
m0}$ in the $\Lambda$CDM model, respectively. The mass of the DM
particle can also be related to the range of $\kappa C_{1}$
(Basilakos \& Plionis 2009)
\begin{equation}
M_x=\frac{1.205 \times 10^{-12}}{\kappa C_{1}} \frac{\langle \sigma
u
  \rangle}{10^{-22}} \; {\rm GeV} \;.
  \label{16}
\end{equation}

 \subsection{Model 2:  Mimicking the $w$CDM Model}
When the annihilation term is negligible ($\kappa=0$) and the
particle creation term dominates, it is straightforward to obtain
the evolution of the total energy density (Basilakos \& Plionis
2009)
  \begin{equation}
\rho(a)=C_{1} a^{n}+(1/C_{2})a^{-3}\;.
      \label{21}
\end{equation}
The conditions $n>-2$ implies that the IDM fluid has an inflection
point, and the current model acts as the $w$CDM model with a
constant EoS with $w_{\rm{IDM}}= -1-n/3$, but with a totally
different intrinsic nature. This situation is mathematically
equivalent to the gravitational DM particle creation process within
the context of non-equilibrium thermodynamics (Lima et al. 2008).
The Hubble parameter is now given by
\begin{equation}
\left(\frac{H}{H_0}\right)^2 =  \Omega_{1,0} a^{n}+\Omega_{\rm 2,0}
a^{-3} \;, \label{22}
\end{equation}
where $\Omega_{\rm 2,0}= 8 \pi G (C_{2}-C_{1})/3H_0^2$ and
$\Omega_{1,0}=8 \pi G C_{1}/ 3H^2_0$, respectively  (Basilakos \&
Plionis 2009).

\section{Observational data}\label{sec3}
To constrain the relevant IDM models (Basilakos \& Plionis 2009), we
use the newly revised $H(z)$ data  (Stern et al. 2010;
Gazta$\tilde{n}$aga et al. 2009) and the Union2 set of 557 SNe Ia
(Amanullah et al. 2010).

\subsection{The observational $H(z)$ data}
It is known that the Hubble parameter $H(z)$ depends on the
differential age as a function of redshift $z$ in the form
\begin{equation}
H(z)=-\frac{1}{1+z}\frac{dz}{dt},
\end{equation}
which provides a direct measurement on $H(z)$ based on $dz/dt$.
Jimenez et al. (2003) demonstrated the feasibility of this method by
applying it to a $z\sim 0$ sample. Moreover, compared with other
observational data, it is more rewarding to investigate the
observational $H(z)$ data directly, because it can take the fine
structure of $H(z)$ into consideration and then use the important
information that this structure provides.

By using the differential ages of passively evolving galaxies
determined from the Gemini Deep Deep Survey (GDDS) (Abraham et al.
2004) and archival data (Treu et al. 2001, 2002; Nolan et al. 2003a,
2003b), Simon et al. (2005) determined nine values of the Hubble
parameter $H(z)$ in the range $0\leq z \leq 1.8$, which have been
used to constrain the parameters of various dark energy models
(Samushia \& Ratra 2006; Wei \& Zhang 2007; Wu \& Yu 2007a, 2007b;
Zhang \& Zhu 2007; Kurek \& Szydlowski 2007; Lazkoz \& Majerotto
2007; Sen \& Scherrer 2008;  Yi \& Zhang 2007; Wan et al. 2007;  Xu
et al. 2008; Zhai, Wan, \& Zhang 2010). The $H(z)$ data at 11
different redshifts were determined from the differential ages of
red-envelope galaxies (Stern et al. 2010), and two more Hubble
parameter data points $H(z=0.24)=79.69\pm4.61$ and
$H(z=0.43)=86.45\pm5.96$ were obtained by Gazta\~{n}aga et al.
(2009) from observations of BAO peaks [for a review of the
observational $H(z)$ data, see Zhang \& Ma (2010)]. Studies using
these newly $H(z)$ data for cosmological constraint include Gong et
al. (2010), Liang et al. (2010a), Liang \& Zhu (2010), Cao \& Liang
(2010), Ma \& Zhang (2011), and Xu \& Wang (2010c). We emphasize the
use of the Hubble constant $H_0$ in our analysis. Many previous have
determined its present value. Freedman et al. (2001) presented  the
final results of the Hubble Space Telescope (\textit{HST}) key
project that measured the Hubble constant $H_0=72\pm8
\rm{kms}^{-1}\rm{Mpc}^{-1}$, Gott et al. (2001) and Chen et al.
(2003) proposed that $H_0=68 \rm{kms}^{-1}\rm{Mpc}^{-1}$ was a more
likely value, and Tammann et al.(2008) obtained $H_0=62.3\pm1.3
\rm{kms}^{-1}\rm{Mpc}^{-1}$ from 28 independently calibrated
Cepheids and the distant, Cepheid-calibrated SNe Ia. More recently,
Riess et al. (2009) determined $H_0=74.2\pm3.6
\rm{kms}^{-1}\rm{Mpc}^{-1}$ by combining the observations of 240
Galactic Cepheid variables using \textit{HST}. In this work,  we
follow Basilakos \& Plionis (2009) and adopt $H_0=72\pm8
\rm{kms}^{-1}\rm{Mpc}^{-1}$. The observational $H(z)$ data are given
in Table~\ref{tab1}. In this case, $\chi^2$ can be defined as
\begin{equation} \chi_{H}^{2} = \sum_i^{14}\frac {(H-H_{\rm obs})^2}{\sigma^2_H}.
\end{equation}

%==================== table 1 ====================
\begin{table*}[htbp]
\begin{center}
\begin{tabular}{c|lllllllllllllllll}\hline\hline
 $z$   &\ 0 & 0.1 & 0.17 & 0.24 & 0.27 & 0.4 & 0.43
       & 0.48 & 0.88 & 0.9 & 1.3 & 1.43 & 1.53 & 1.75\\ \hline
 $H(z)$ &\ 72 & 69 & 83 & 79.69 & 77 & 95 & 86.45
       & 97 & 90 & 117 & 168 & 177 & 140 & 202\\ \hline
 $1 \sigma$ uncertainty &\ $\pm 8$ & $\pm 12$ & $\pm 8$ & $\pm 4.61$ & $\pm 14$ & $\pm 17$ & $\pm 5.96 $
  & $\pm 60$ & $\pm 40$ & $\pm 23$ & $\pm 17$ & $\pm 18$ & $\pm 14$ & $\pm 40$ \\ \hline\hline

\end{tabular}
\end{center}
\caption{\label{tab1} The observational $H(z)$ data (Stern et al.
2010; Gazta\~{n}aga et al. 2009; Freedman et al. 2001)}
\end{table*}
%=================================================
\subsection{The observational SNe Ia data}

SNe Ia have long been used as ``standard candles''. It is commonly
believed that measuring both their redshifts and apparent peak
fluxes gives a direct measurement of their luminosity distances,
thus SNe Ia provide the strongest constraints on the cosmological
parameters (Riess et al. 2004, 2007; Astier et al. 2006; Davis et
al. 2007; Wood-Vasey et al. 2007; Kowalski et al. 2008; Hicken et
al. 2009; Chen et al. 2010). The present analysis uses the Union2
(557 sample) data set of the Supernova Cosmology project covering a
redshift range $0.015 \leq z \leq 1.4$ (Amanullah et al. 2010),
which was used to constrain cosmological models in Wei (2010), Liang
et al. (2010a, 2010b), Liang et al. (2011), Xu \& Wang (2010a,
2010b), and Wang et al. (2010).

In the calculation of the likelihood from SNe Ia, we marginalize the
nuisance parameter by minimizing (Di Pietro \& Claeskens 2003),
\begin{equation}
\chi^2_{\rm SNe}=A-\frac{B^2}{C}+\ln\left(\frac{C}{2\pi}\right),
\end{equation}
where $A=\sum_i^{557}{(\mu^{\rm data}-\mu^{\rm
th})^2}/{\sigma^2_i}$, $B=\sum_i^{557}{(\mu^{\rm data}-\mu^{\rm
th})}/{\sigma^2_i}$, $C=\sum_i^{557}{1}/{\sigma^2_i}$, and the
distance modulus is $\mu=5 \log(d_L/\rm{Mpc})+25$, with the
$1\sigma$  uncertainty $\sigma_i$ from the observations of SNe Ia
and the luminosity distance $d_L$ as a function of redshift $z$,
\begin{equation}
d_L=(1+z)\int^{z}_0\frac{cdz'}{H(z')}~.
\end{equation}

\section{Constraint results} \label{sec4}
We define the total likelihood to be the product of the separate
likelihoods of the two cosmological probes, in other words,
\begin{equation} \chi_{\rm total}^{2}=\chi_{H}^{2} + \chi_{\rm SNe}^{2}.
\end{equation}
To determine the best-fit cosmological parameters, we use a $\chi^2$
minimization and the 68.3\% and 95.4\% confidence levels are defined
by the conventional two-parameters $\chi^2$ levels 2.30 and 6.17,
respectively. We perform a global fitting to determine the
cosmological parameters using the Markov chain Monte Carlo (MCMC)
method and our MCMC code is based on the publicly available package
COSMOMC (Lewis \& Bridle 2002).

\subsection{Model 1:  Mimicking the $\Lambda$CDM Model}

\begin{figure}
\begin{center}
\includegraphics[width=0.45\hsize]{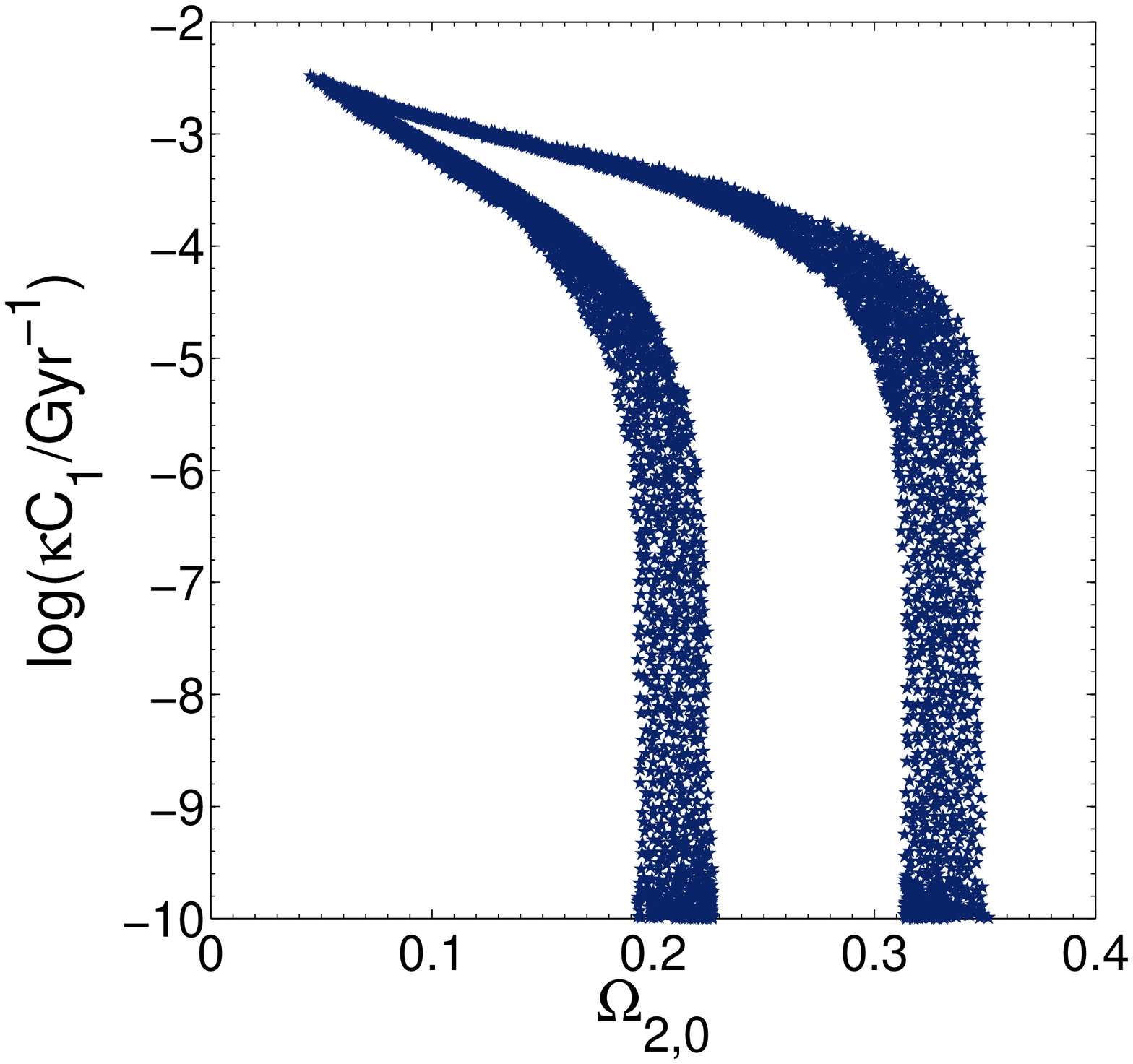}
\includegraphics[width=0.45\hsize]{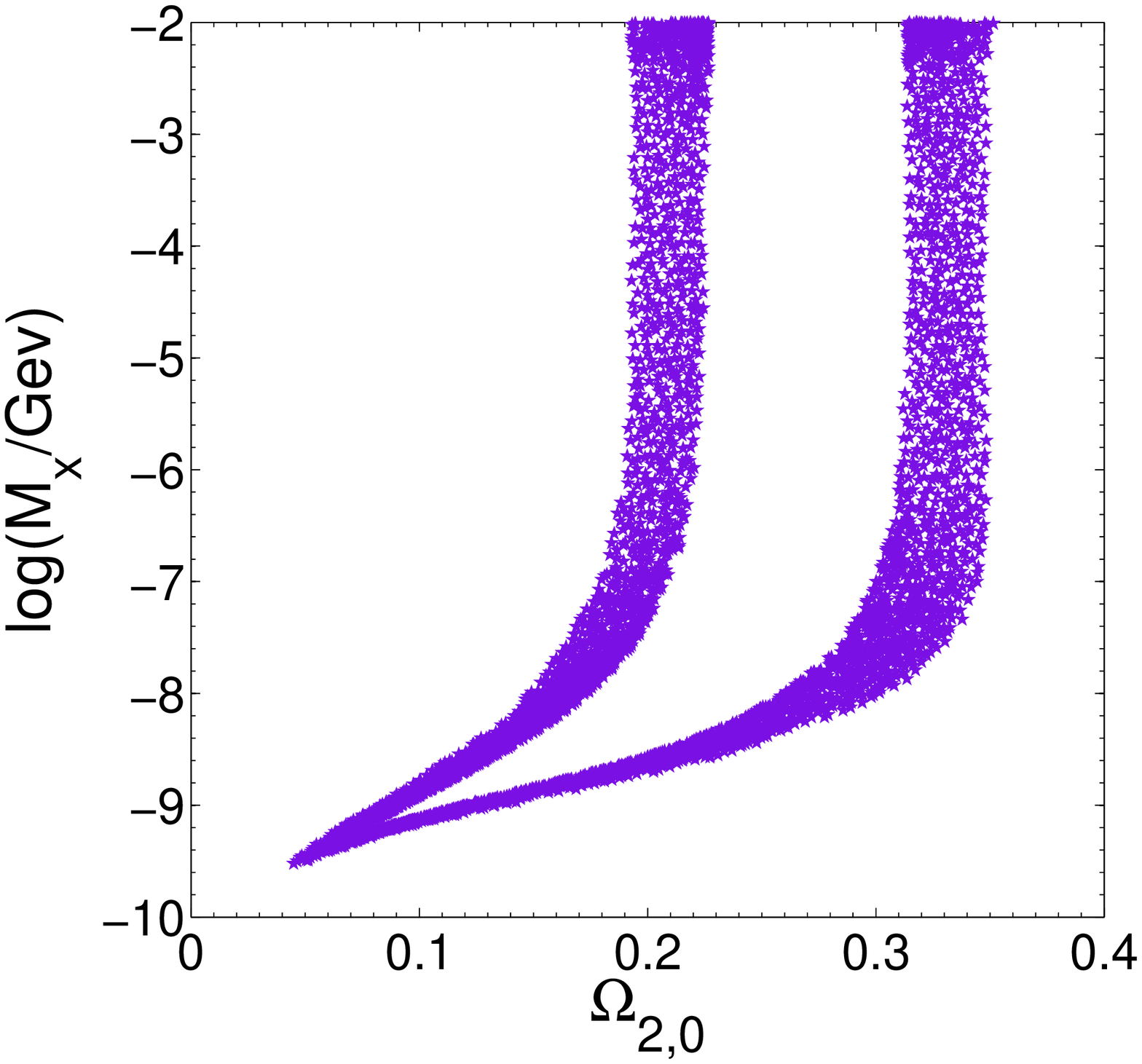}
\end{center}
\caption{The likelihood contours at the 68.3\% and 95.4\% confidence
levels in the $\Omega_{2,0}-\kappa C_1$ and $\Omega_{2,0}-M_x$
planes provided by fitting model 1 to the  $H(z)$ data.
\label{fig1.1}}
\end{figure}

\begin{figure}
\begin{center}
\includegraphics[width=0.45\hsize]{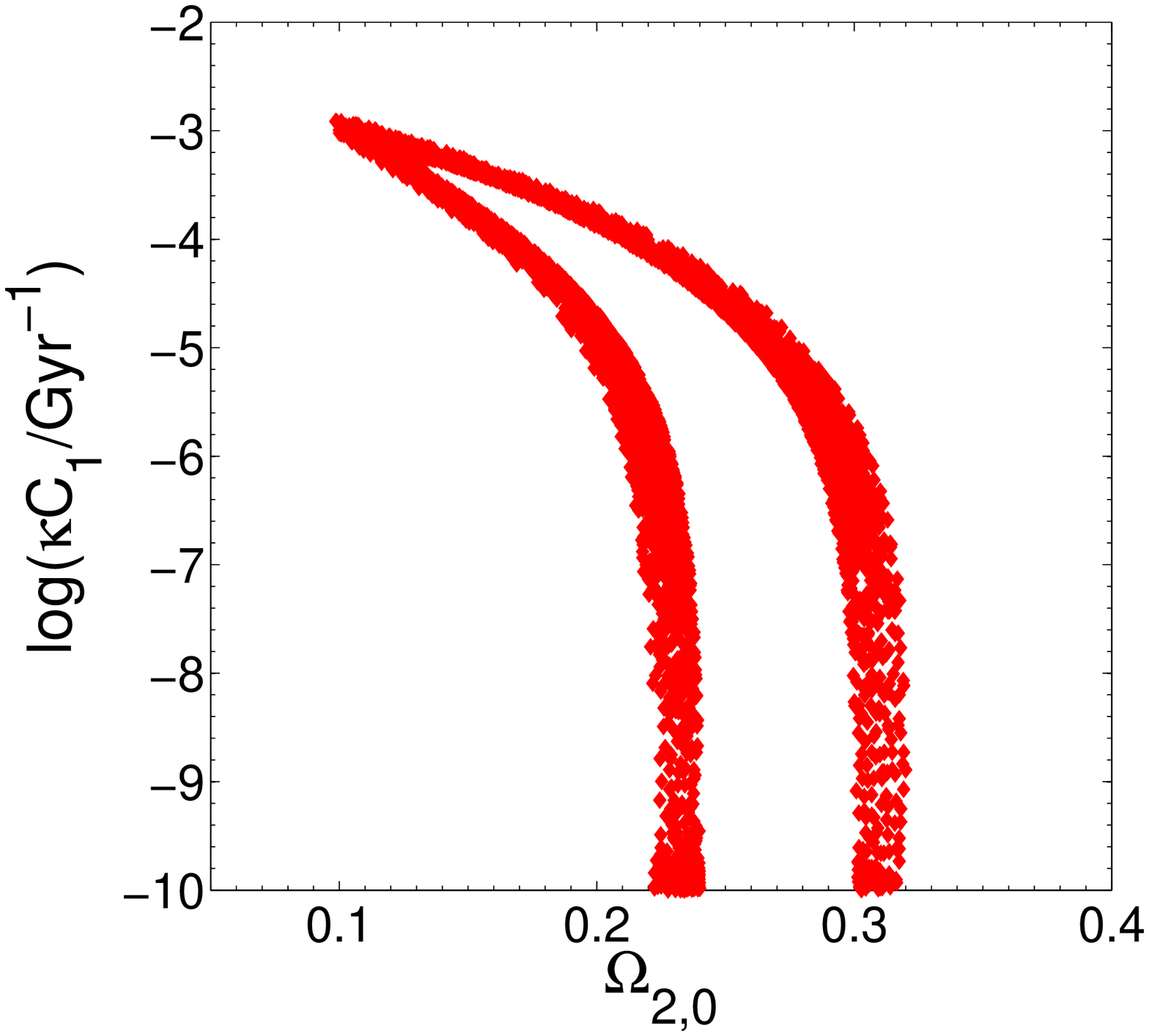}
\includegraphics[width=0.45\hsize]{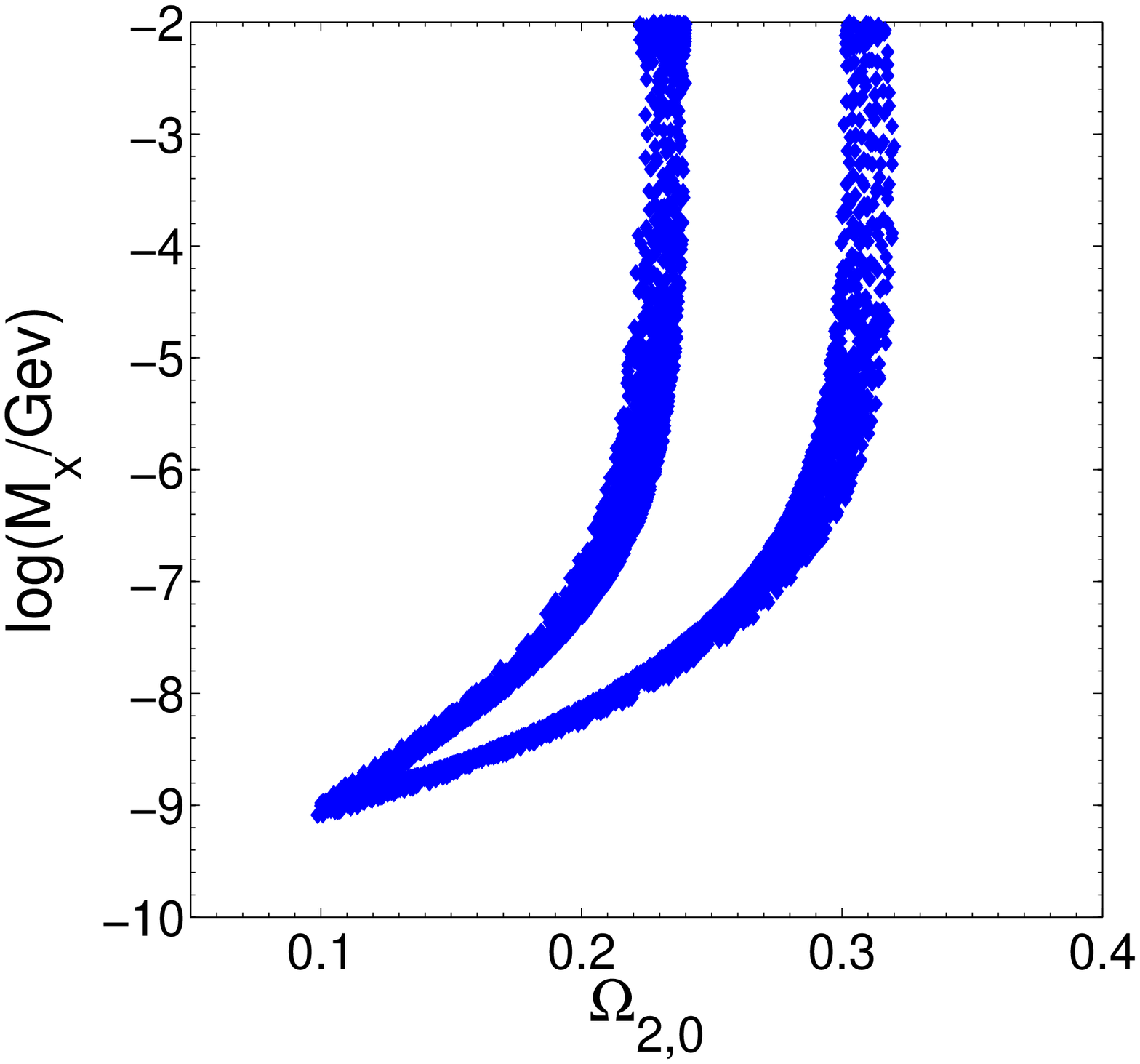}
\end{center}
\caption{The likelihood contours at the 68.3\% and 95.4\% confidence
levels in the $\Omega_{2,0}-\kappa C_1$ and $\Omega_{2,0}-M_x$
planes provided by fitting model 1 to the SNe Ia data.
\label{fig1.2}}
\end{figure}

\begin{figure}
\begin{center}
\includegraphics[width=0.45\hsize]{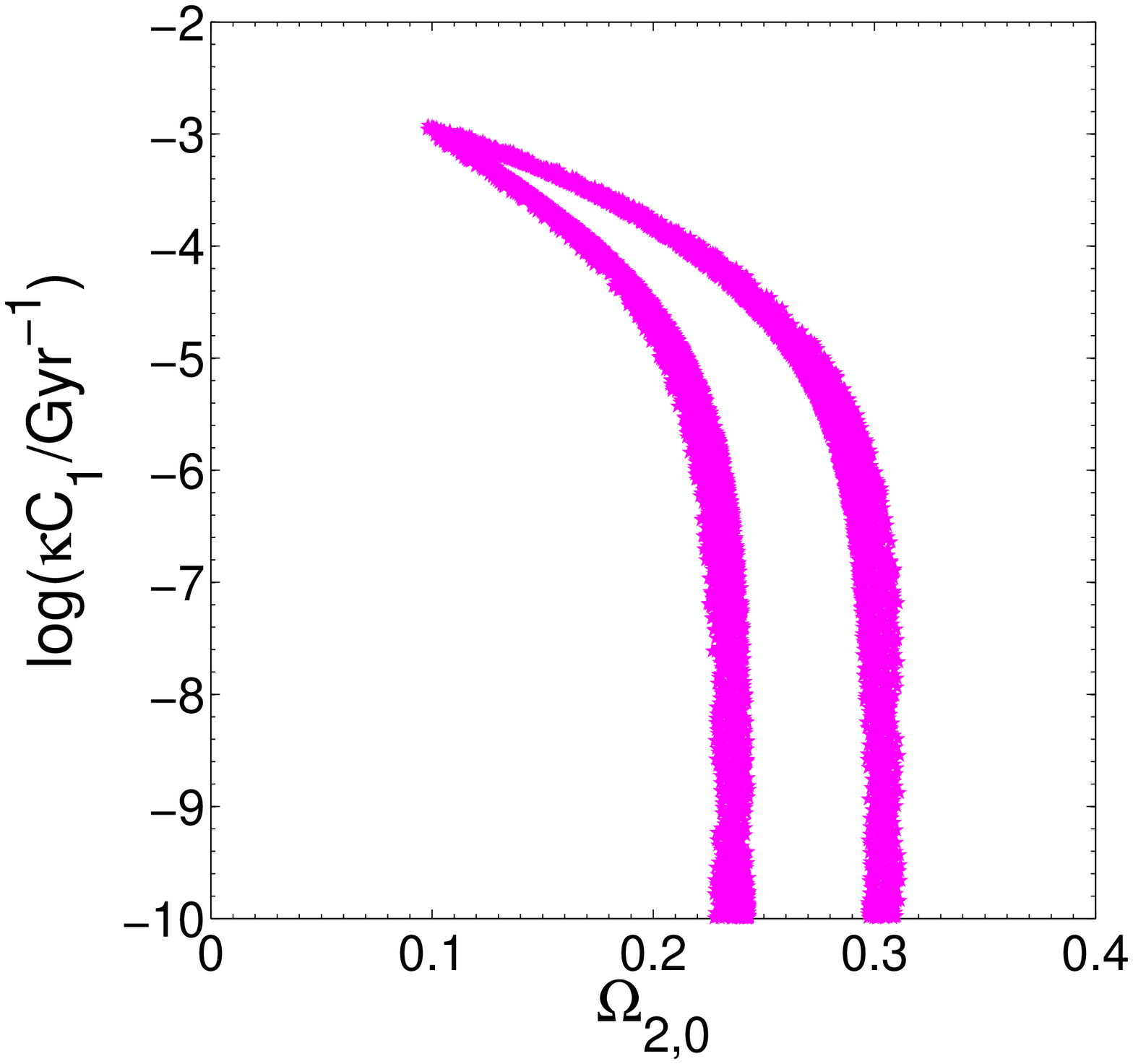}
\includegraphics[width=0.45\hsize]{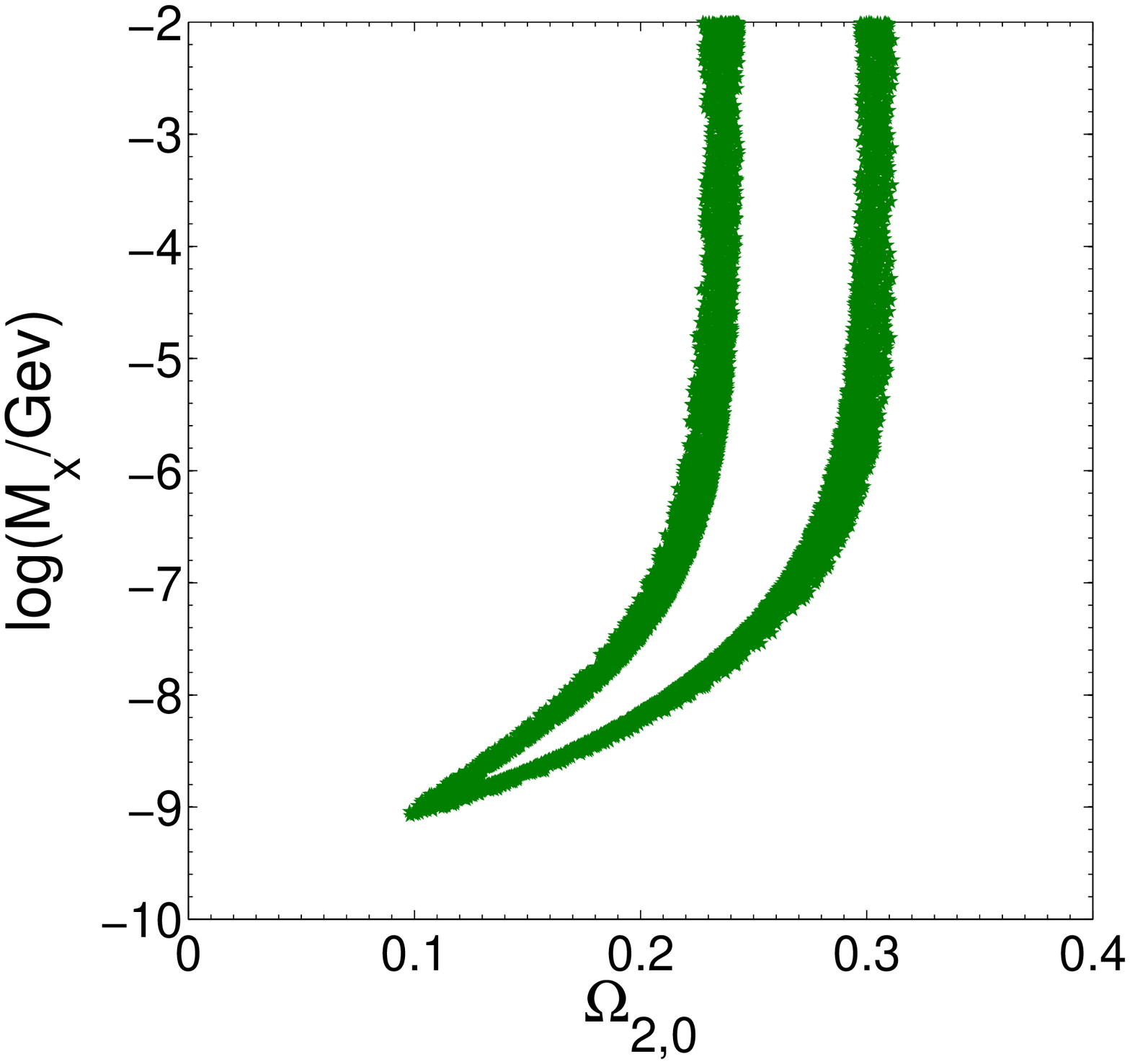}
\end{center}
\caption{The likelihood contours at the 68.3\% and 95.4\% confidence
levels in the $\Omega_{2,0}-\kappa C_1$ and $\Omega_{2,0}-M_x$
planes provided by fitting model 1 to the $H(z)$+SNe data. \label{fig1.3}}
\end{figure}

In this model, there are two free parameters: $\Omega_{2,0}$ and
$\kappa C_1$ (or $M_x$) for a flat background
($\Omega_{1,0}+\Omega_{2,0}=1$). In this paper, we follow Basilakos
\& Plionis (2009) and adopt $t_0 =1/ H_0 =13.6\ \textrm{Gyrs}$
(roughly the age of the universe in the corresponding $\Lambda$CDM
cosmology). The constraint results from different data combinations
are shown in Fig.~\ref{fig1.1} - \ref{fig1.3} and summarized in
Table~\ref{tab2}.

\begin{table}[htbp]
\begin{center}
\begin{tabular}{c|c|c|c}\hline\hline
 Model 1 & $H(z)$
 & SNe
 & $H(z)$+SNe\\ \hline
 $\ \chi_{\rm min}^2\ $ & 9.86 & 544.13 & 554.34\\
 $\Omega_{2,0}$ & $0.270^{+0.044}_{-0.044}$& $0.271^{+0.033}_{-0.031}$ & $0.272^{+0.028}_{-0.029}$\\
$\log(\kappa C_1\cdot\textrm{Gyr})$ & $-6.85 (\le-3$)& $-4.42
(\le-3.5)$
& $-5.35 (\le-3.4)$ \\

$\log M_x/\textrm{Gev}$ & $-5.15 (\ge-9)$ & $-7.58 (\ge-8.5)$ & $-6.65 (\ge-8.6)$ \\

 \hline\hline
\end{tabular}
\end{center}
\caption{\label{tab2} Summarizing the results of parameter
constraints from model 1 considered in this work. We adopt $t_0 =1/
H_0 =13.6 \rm Gyrs$ in the computations.}
\end{table}

%\clearpage

In Fig.~\ref{fig1.1}, we show the results from the $H(z)$ data. The
best-fit solution corresponds to fit
$\Omega_{2,0}=0.270^{+0.044}_{-0.044}$ and $\log(\kappa
C_1\cdot\textrm{Gyr})\approx -6.85$ with an upper limit $\log(\kappa
C_1\cdot\textrm{Gyr})\approx -3$. This effective annihilation term
is obviously still unconstrained towards lower values.
Correspondingly, the best-fit value of $M_x$ is $\log
M_x/\textrm{Gev}\approx -5.15$ with a relatively stringent lower
limit of $\log M_x/\textrm{Gev}\approx -9$. These results are
consistent with previous work that uses fewer observational $H(z)$
data $\Omega_{2,0}=0.3^{+0.05}_{-0.08}$ and $\log(\kappa
C_1\cdot\textrm{Gyr})\approx -9.3$ (Basilakos \& Plionis 2009). In
Fig.~\ref{fig1.2}, we show the constraint results from SNe Ia with
the best-fit model parameters $\Omega_{2,0}=0.271^{+0.033}_{-0.031}$
and $\log(\kappa C_1\cdot\textrm{Gyr})\approx -4.42$. A more
stringent upper limit is obtained at around $\log(\kappa
C_1\cdot\textrm{Gyr})\approx-3.5$. The best-fit DM particle mass is
$\log M_x/\textrm{Gev}\approx -7.58$ with a relatively stringent
lower limit $\log M_x/\textrm{Gev} \approx -8.5$. To obtain a
tighter constraint on the model parameters, we combine the $H(z)$
and SNe Ia data, and the results are shown in Fig.~\ref{fig1.3}. The
best-fit model parameters are $\Omega_{2,0}=0.272^{+0.028}_{-0.029}$
and $\log(\kappa C_1\cdot\textrm{Gyr})\approx -5.35$ with a much
more stringent upper limit $\log(\kappa C_1\cdot\textrm{Gyr})\approx
-3.4$. Moreover, the best-fit value of $M_x$ is $\log
M_x/\textrm{Gev}\approx -6.65$ with a relatively stringent lower
limit of $\log M_x/\textrm{Gev}\approx -8.6$ at $2\sigma$. Since
$M_x$ is unbound at small values, it is consistent with currently
accepted lower bounds of $M_x$ (10GeV) (see Cirelli et al.(2009) and
references therein).

\subsection {Model 2: Mimicking the $w$CDM model}

\begin{figure}
\begin{center}
\includegraphics[width=0.45\hsize]{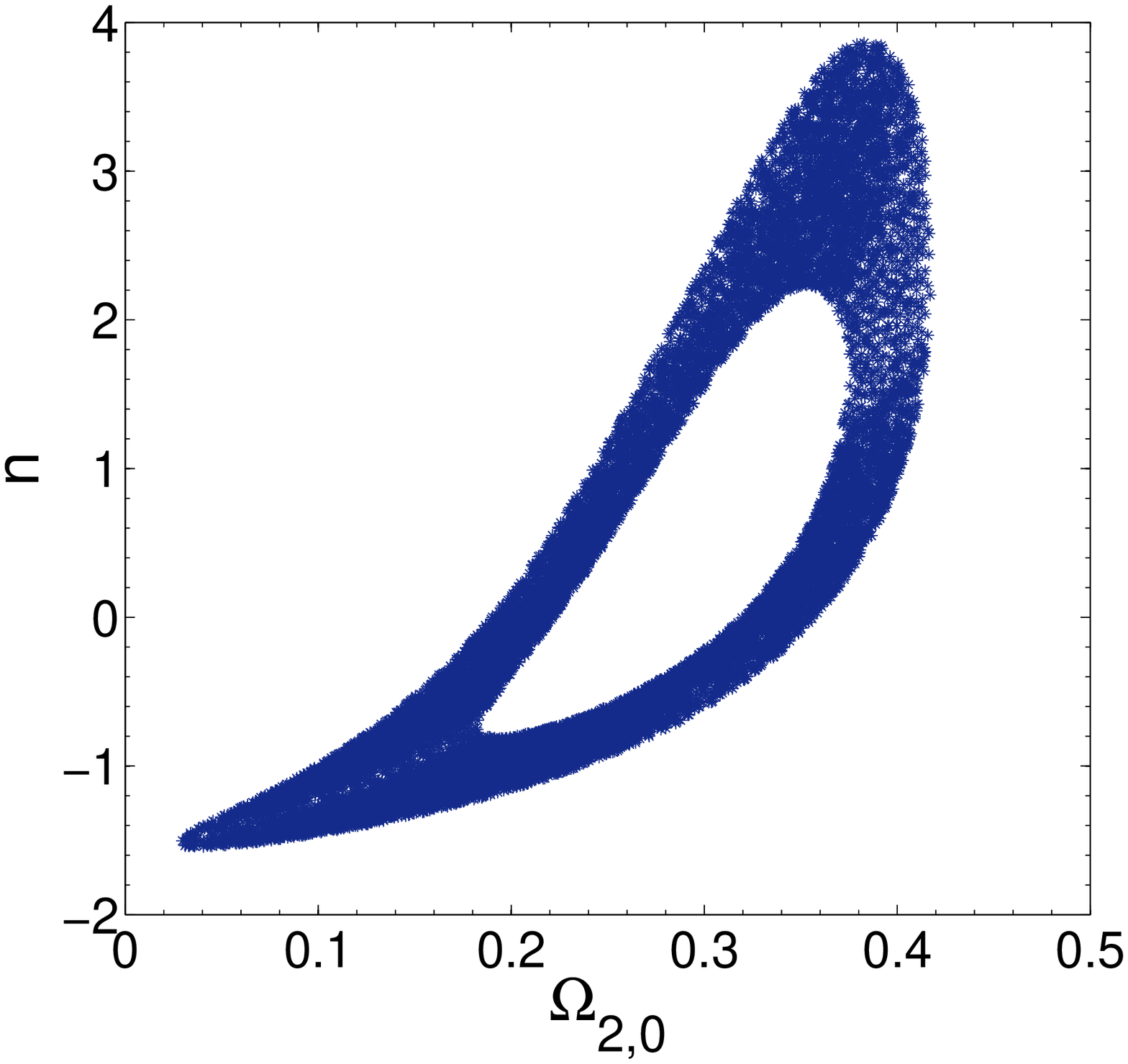}
\includegraphics[width=0.45\hsize]{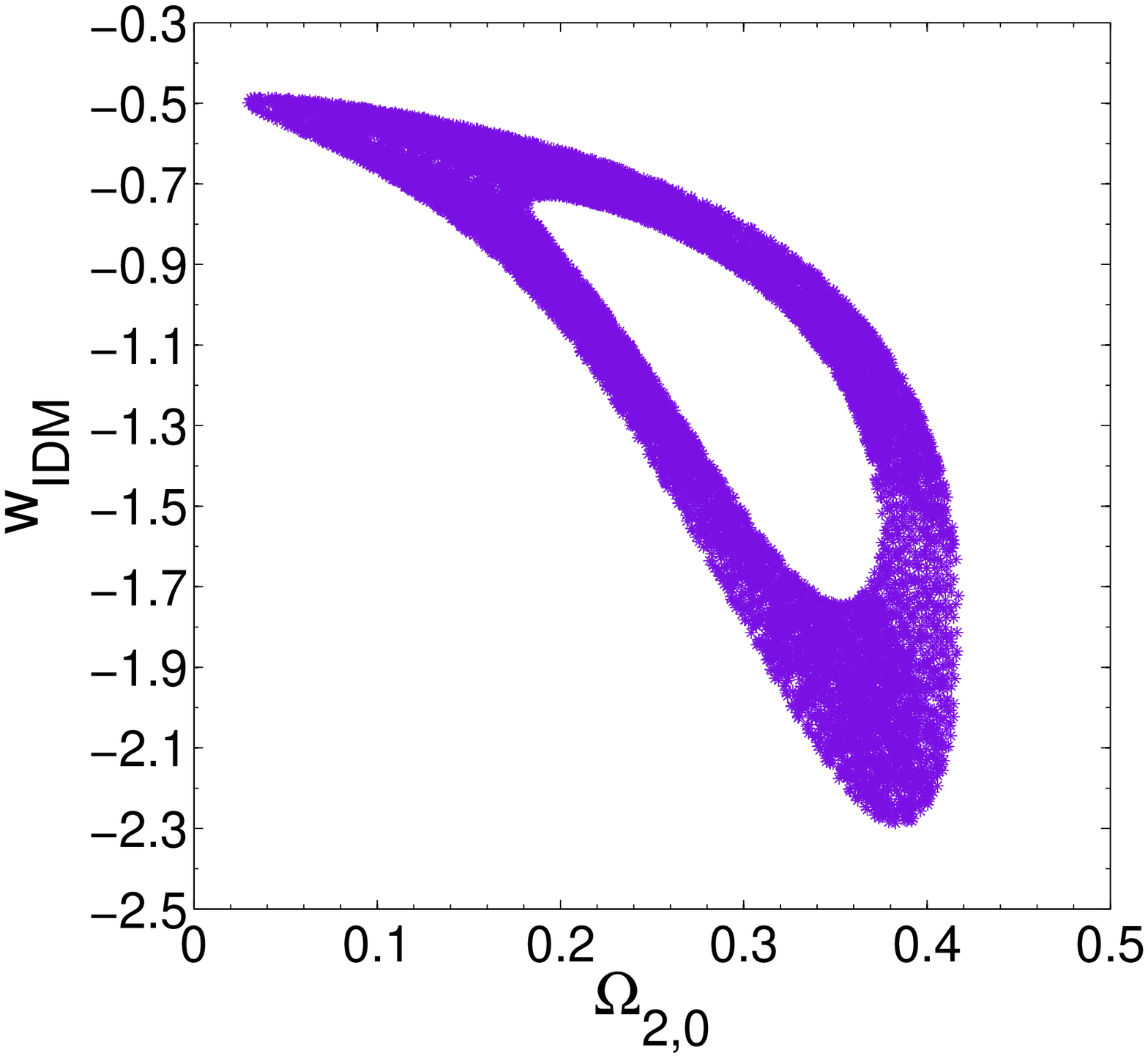}
\end{center}
\caption{The likelihood contours at the 68.3\% and 95.4\% confidence
levels in the $\Omega_{2,0}-n$ and $\Omega_{2,0}-w_{\rm IDM}$ planes
provided by fitting model 2 to the $H(z)$ data. \label{fig2.1}}
\end{figure}

\begin{figure}
\begin{center}
\includegraphics[width=0.45\hsize]{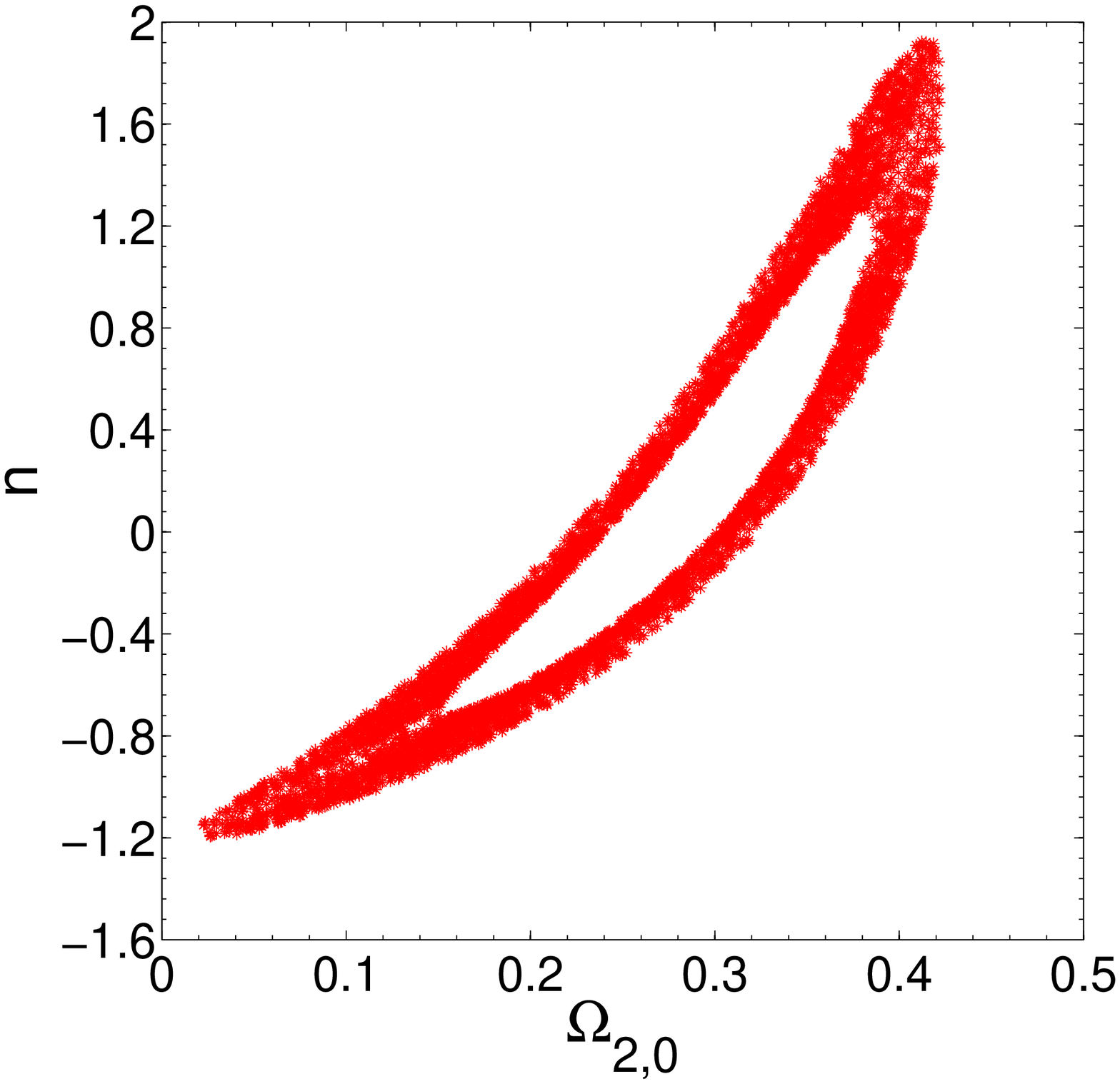}
\includegraphics[width=0.45\hsize]{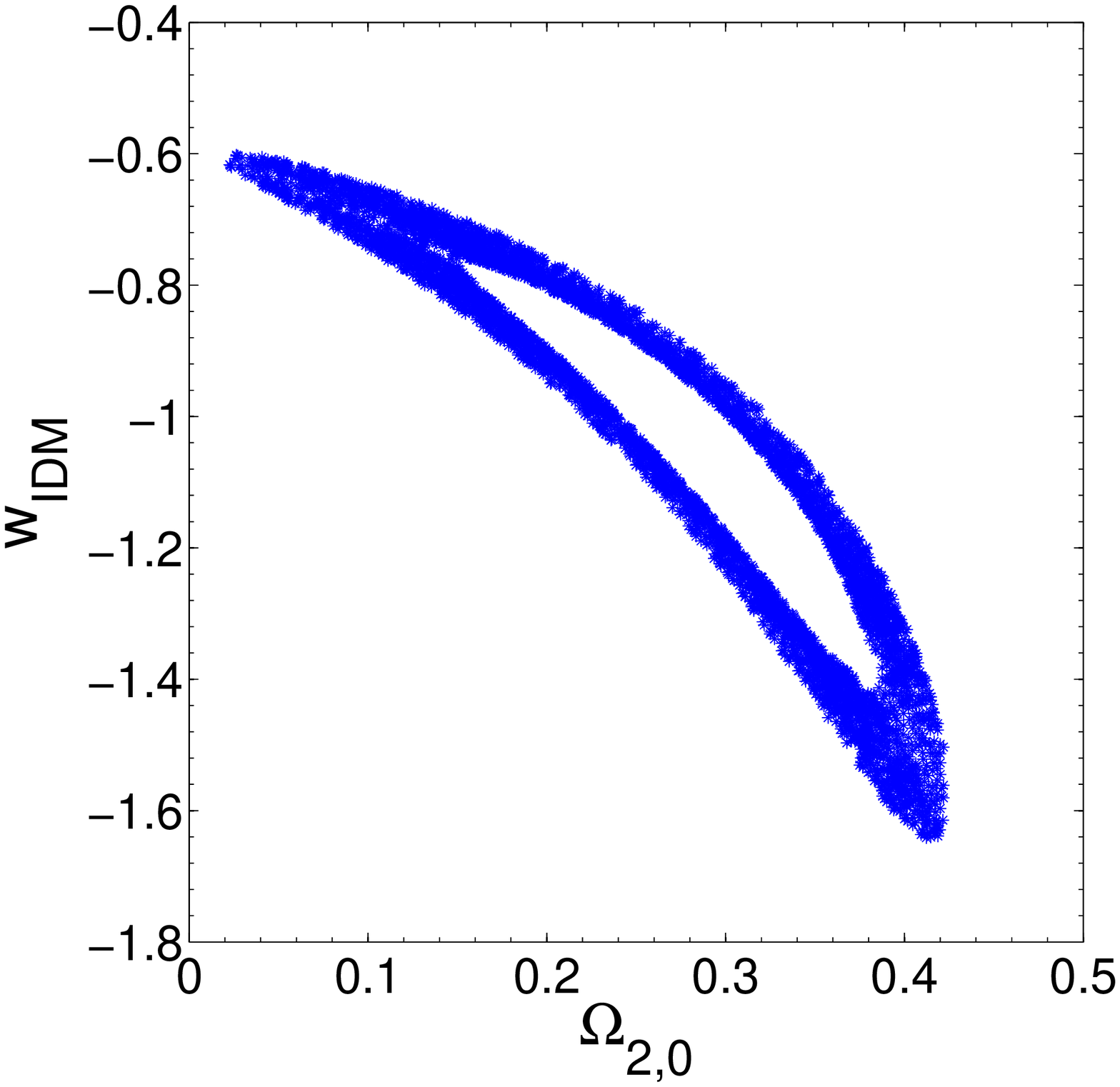}
\end{center}
\caption{The likelihood contours at the 68.3\% and 95.4\% confidence
levels in the $\Omega_{2,0}-n$ and $\Omega_{2,0}-w_{\rm IDM}$ planes
provided by fitting model 2 to the SNe Ia data. \label{fig2.2}}
\end{figure}

\begin{figure}
\begin{center}
\includegraphics[width=0.45\hsize]{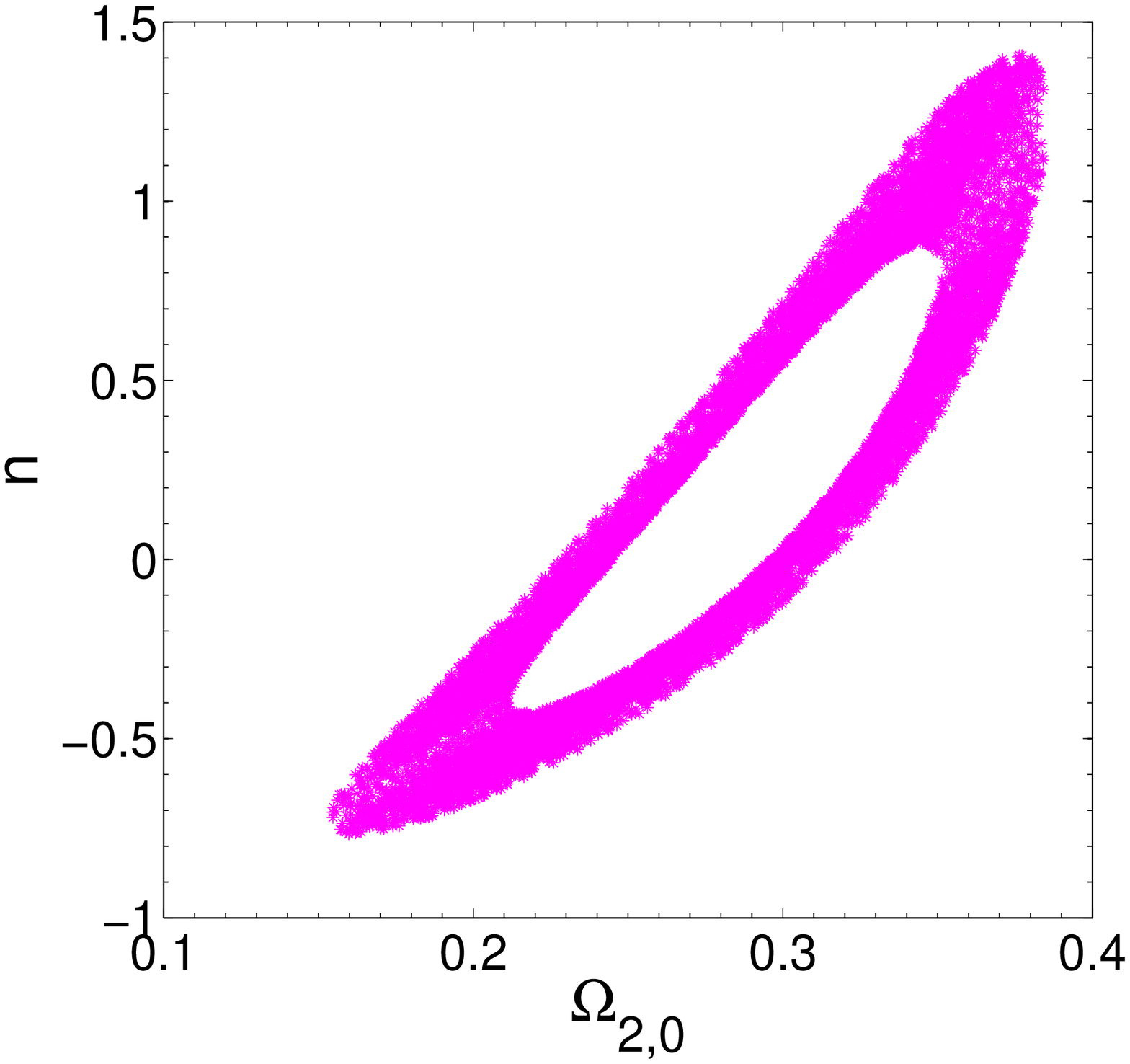}
\includegraphics[width=0.45\hsize]{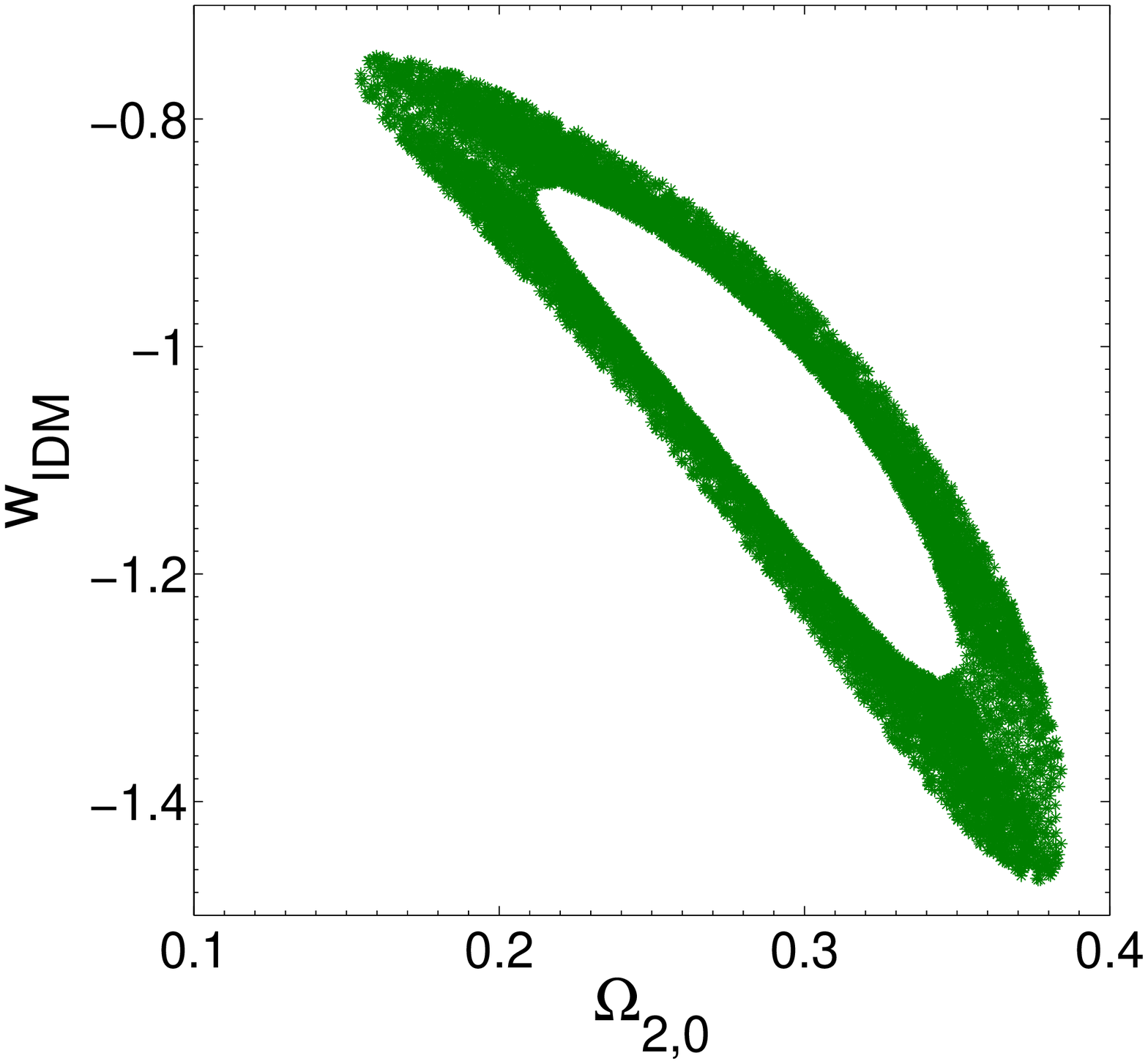}
\end{center}
\caption{The likelihood contours at the 68.3\% and 95.4\% confidence
levels in the $\Omega_{2,0}-n$ and $\Omega_{2,0}-w_{\rm IDM}$ planes
provided by fitting model 2 to the $H(z)$+SNe data.
 \label{fig2.3}}
\end{figure}

In this case, there are two free parameters: $\Omega_{2,0}$ and $n$
(or $w_{\rm{IDM}}$). The constraint results from different data
combinations are shown in Fig.~\ref{fig2.1} - Fig.~\ref{fig2.3} and
summarized in Table~\ref{tab3}.

\begin{table}[htbp]
\begin{center}
\begin{tabular}{c|c|c|c}\hline\hline
 Model 2 & $H(z)$
 &SNe
 & $H(z)$+SNe\\ \hline
 $\ \chi_{\rm min}^2\ $ & 9.40 & 544.12 & 554.39\\
 $\Omega_{2,0}$         & $0.277^{+0.098}_{-0.097}$& $0.288^{+0.108}_{-0.105}$ & $0.281^{+0.072}_{-0.071}$ \\
 $n$                    & $0.70^{+1.54}_{-1.54}$& $0.17^{+1.10}_{-1.10}$ & $0.22^{+0.67}_{-0.66}$ \\
$w_{\rm{IDM}}$      & $-1.23^{+0.51}_{-0.51}$ & $-1.06^{+0.37}_{-0.37}$ & $-1.07^{+0.24}_{-0.23}$ \\

 \hline\hline
\end{tabular}
\end{center}
\caption{\label{tab3} Summarizing the results of constraint on
parameters from model 2.}
\end{table}

For the $H(z)$ data and the results shown in Fig.~\ref{fig2.1}, the
best-fit parameter values are $\Omega_{2,0}=0.277^{+0.098}_{-0.097}$
and $n=0.70^{+1.54}_{-1.54}$. For the EoS of IDM, the best-fit value
is $w_{\rm{IDM}}=-1.23^{+0.51}_{-0.51}$. These results are
consistent with the previous results of Basilakos \& Plionis (2009)
by using fewer observational $H(z)$ data ($n\simeq -0.30$ and
$w_{\rm{IDM}}\simeq -0.90$, which favors the effective quintessence
model with a constant EoS for which $w>-1$). In Fig.~\ref{fig2.2},
we show the constraint results from SNe Ia. By minimizing the
corresponding $\chi^2$, we find that the best-fit model values are
$\Omega_{2,0}=0.288^{+0.108}_{-0.105}$ and $n= 0.17^{+1.10}_{-1.10}$
($w_{\rm{IDM}}= -1.06^{+0.37}_{-0.37}$). For the combined data
$H(z)$+ SNe shown in Fig.~\ref{fig2.3}, we find the best-fit values
are $\Omega_{2,0}=0.281^{+0.072}_{-0.071}$ and
$n=0.22^{+0.67}_{-0.66}$ ($w_{\rm{IDM}}=-1.07^{+0.24}_{-0.23}$).
Obviously, all of the above constraints are consistent with the
concordance $\Lambda$CDM model and appear to be most consistent with
the effective phantom model and a constant EoS for which $w<-1$.

\section{Conclusions} \label{sec5}

We have investigated the interacting dark matter (IDM) scenario
mimicking either the $\Lambda$CDM model or the $w$CDM model, which
can create the cosmic acceleration without dark energy (Basilakos \&
Plionis 2009). In our work, the scale of the effective annihilation
term $\kappa C_{1}$ and therefore the mass of DE particles $M_x$ are
constrained with different newly revised observational data
including $H(z)$ and Union2 SNe Ia data. When mimicking a $\Lambda$
cosmology and using three different data combinations of $H(z)$, SNe
Ia, and SNe Ia+$H(z)$, we have found that $\kappa C_1$ is quite
small, which is consistent with the previous results in Basilakos \&
Plionis (2009), which used fewer observational $H(z)$ data.
Meanwhile, for the combined data sets, we obtain a more stringent
upper limit to the effective annihilation term with $\log(\kappa
C_1\cdot\textrm{Gyr})\approx -3.4$. By relating the range of values
of $\kappa C_1$ to the mass of the DM particle, we have inferred an
apparent lower limit of $M_x\approx 10^{-8.6}\textrm{Gev}$.
Furthermore, when mimicking $w$CDM model and assuming that the
particle creation term dominates ($\kappa=0$), we obtained the
effective equation of state of IDM is consistent with the
concordance $\Lambda$CDM model and appears to be consistent with the
effective phantom model with a constant EoS for which $w<-1$.
%which is inconsistent with the
%previous results in Basilakos \& Plionis (2009).

To sum up, we conclude that the interacting dark matter (IDM) model
may provide a practical alternative to Dark Energy in explaining the
present cosmic acceleration. The hope of proving this conclusion
should be pinned on analyses of future observational data such as
high redshift SNe Ia data from SNAP (Albrecht et al. 2006), more
precise CMB data from the ESA Planck satellite (Balbi 2007), and
complementary data, such as the X-ray gas mass fraction in clusters
(Allen et al. 2004; Allen et al. 2008; Ettori et al. 2009),
gravitational lensing data (Zhu 1998; Sereno 2002), as well as
gamma-ray bursts (GRBs) at high redshift  (Schaefer 2007; Liang et
al. 2008; Basilakos \& Perivolaropoulos 2008; Liang \& Zhang 2008;
Wang \& Liang 2010;  Gao et al. 2010; Liang et al. 2011).

%__________________________________________________________________
\begin{acknowledgements}
We thank Yun Chen, Hao Wang, Yan Dai, Chunhua Mao, Fang Huang, Yu
Pan, Jing Ming, Kai Liao and Dr. Yi Zhang for discussions. This work
was supported by the National Science Foundation of China under the
Distinguished Young Scholar Grant 10825313, the Key Project Grants
10533010, and by the Ministry of Science and Technology national
basic science Program (Project 973) under grant No. 2007CB815401.
\end{acknowledgements}

%\clearpage


\begin{thebibliography}{}

\bibitem [Abraham et al.(2004)]{Abraham04} Abraham, R. G., et al. 2004, ApJ, 127, 2455
\bibitem [Albrecht et al.(2006)]{Abraham04} Albrecht, A., et al. 2006, Report of the Dark Energy Task Force, arXiv:0609591
\bibitem [Allen et al.(2004)]{Allen04} Allen, S.~W., et al. 2004, MNRAS, 353, 457
\bibitem [Allen et al.(2008)]{Allen08} Allen, S.~W., et al. 2008, MNRAS, 383, 879
\bibitem [Amanullah et al.(2010)] {Amanullah} Amanullah, R., et al. [Supernova Cosmology Project Collaboration], 2010, ApJ, 716, 712
\bibitem [Astier et al.(2006)]{Astier06} Astier, P., et al. 2006, A\&A, 447, 31
\bibitem [Balakin et al.(2003)]{Balakin03} Balakin, A. B., Pavon, D., Schwarz, D. J., \& Zimdahl, W. 2003, N. J. Phys, 5, 85
\bibitem [Balbi(2007)]{Balbi2007}Balbi, A. 2007, New A. R., 51, 281
\bibitem []{} Basilakos, S., \& Perivolaropoulos, L. 2008, MNRAS, 391, 411
\bibitem [Basilakos \& Plionis(2009)]{Basilakos2009} Basilakos, S., \& Plionis, M. 2009, A\&A, 507, 47 %[arXiv:0807.4590]
\bibitem [Basilakos \& Lima  (2010)] {Basilakos2010} Basilakos,  S., \& Lima, J. A. S. 2010, PRD, 82, 023504 %[arXiv:1003.5754]
\bibitem []{} Bengochea, G. R., \& Ferraro, R. 2009, PRD, 79, 124019
\bibitem [Bento et al.(2002)] {Bento02}Bento, M. C., et al. 2002, PRD, 66, 043507
\bibitem []{} Cao, S., \& Liang, N. 2010, arXiv:1012.487
\bibitem [Caldwell et al.(1998)]{Caldwell1998} Caldwell, R., Dave, R., \& Steinhardt, P. J. 1998, PRL, 80, 1582
\bibitem [Caldwell (2002)]{}  Caldwell, R. 2002, PLB, 545, 23   %phantom
\bibitem []{}  Capozziello, S., \& Fang, L. Z. 2002, International Journal of Modern Physics D, 11, 483    %f(R)
\bibitem[]{}  Carroll S. M., Duvvuri V., Trodden M., \& Turner M. S. 2004, PRD, 70, 043528

\bibitem [Chen et al.(2010)]{Chen10} Chen, Y., Zhu, Z.-H., Alcaniz, J. S., \& Gong, Y. G. 2010, ApJ, 711, 439
\bibitem []{} Chen, G., Gott, J. R., III, \& Ratra, B. 2003, Publ. Astron. Soc. Pac., 115, 1269
\bibitem [Cirelli et al.(2009)]{Cirelli09} Cirelli, M., Iocco, F., \& Panci, P. 2009, JCAP, 10, 9
\bibitem [Cohen et al.(1999)]{Cohen99}Cohen, A., Kaplan, D., \& Nelson, A. 1999, PRL, 82, 4971
\bibitem [Davis et al.(2007)]{Davis07}Davis, T. M., et al. 2007, ApJ, 666, 716
\bibitem [Dvali, Gabadadze \& Porrati 2000]{}  Dvali, G., Gabadadze, G., \& Porrati, M. 2000, PLB, 485, 208
\bibitem [Di Pietro \& Claeskens(2003)]{Pietro03} Di Pietro, E., \& Claeskens, J. F. 2003, MNRAS, 341, 1299
\bibitem [Ettori et al.(2009)]{Ettori09}Ettori, S., et al. 2009, AAP, 501, 61
\bibitem [Farrar \& Peebles(2004)]{Farrar04} Farrar, G. R., \& Peebles, P. J. E. 2004, ApJ, 604, 1
\bibitem [Feng, Wang, X. and Zhang 2005]{} Feng, B., Wang, X., \& Zhang, X. 2005, PLB, 607, 35  %astro-ph/0404224
\bibitem []{} Freedman, W. L., et al. 2001, ApJ,  553, 47
\bibitem []{} Freese, K., \& Lewis, M. 2002, PLB, 540, 1

\bibitem []{} Gao, H., Liang, N., \& Zhu, Z.-H. 2010, arXiv:1003.5755
\bibitem [Gazta\~{n}aga et al.(2009)]{hz3} Gazta\~{n}aga, E., Cabr\'{e}, A., \& Hui, L. 2009, MNRAS, 399, 1663
\bibitem []{} Gong, Y. G., Cai, R. G., Chen, Y., \& Zhu, Z.-H. 2010, JCAP, 01, 019 %[arXiv:0909.0596]
\bibitem []{}Gott, J. R., III, Vogeley, M. S., Podariu, S., \& Ratra, B. 2001, ApJ, 549, 1
\bibitem [Gubser \& Peebles(2004)]{Gubser04} Gubser, S. S., \& Peebles, P. J. E. 2004, PRD, 70, 123510
\bibitem [Guo et al. 2005]{} Guo, Z.-K., Piao, Y.-S., Zhang, X., \& Zhang, Y. Z. 2005, PLB, 608, 177
\bibitem [Hicken et al.(2009)]{Hicken09} Hicken, M., et al. 2009, ApJ, 700, 1097
\bibitem []{} Jassal, H. K., Bagla, J. S., \& Padmanabhan, T. 2010, MNRAS, 405, 2639
\bibitem [Jimenez et al.(2003)]{Jimenez03} Jimenez, R., Verde, L., Treu, T., \& Stern., D. 2003, ApJ, 593, 622
\bibitem [Kamenshchik et al.(2001)] {Kamenshchik01}Kamenshchik, A., Moschella, U., \& Pasquier, V. 2001, PLB, 511, 265
\bibitem [Kolb \& Turner(1990)]{Kolb90} Kolb, E. W, \& Turner, M. S. 1990, The Early Universe, Addison-Wesley Publishing
\bibitem [Kowalski et al.(2008)]{Kowalski08} Kowalski, M., et al. 2008, ApJ, 686, 749
\bibitem [Kurek \& Szydlowski(2007)]{Kurek07} Kurek, A., \& Szydlowski, M. 2008, ApJ, 675, 1
\bibitem [Lazkoz \& Majerotto(2007)]{Lazkoz07} Lazkoz, R., \& Majerotto, E. 2007, JCAP, 0707, 015
\bibitem [Lewis \& Bridle(2002)] {Lewis02} Lewis, A., \& Bridle, S. 2002, PRD, 66, 103511
\bibitem [Li(2004)]{Li04} Li, M. 2004, PLB, 603, 1
\bibitem [Liang et al.(2008)]{Liang2008} Liang, N., Xiao, W. K., Liu, Y., \& Zhang, S. N. 2008, ApJ, 685, 354
\bibitem []{}    Liang, N. \& Zhang, S. N. 2008, AIP Conf. Proc., 1065, 367
\bibitem [Liang et al.(2009)]{Liang2009} Liang, N., Gao, C. J., \& Zhang, S. N., Chin. Phys. Lett., 2009, 26, 069501 %[arXiv:0904.4626]
\bibitem [Liang Wu \& Zhang(2010)]{Liang2010} Liang, N., Wu, P., \& Zhang, S. N. 2010a, PRD, 81, 083518 %[arXiv:0911.5644]
\bibitem [Liang Wu \& Zhu(2010)]{}    Liang, N., Wu, P., \& Zhu, Z.-H. 2010b, arXiv:1006.1105
\bibitem [Liang Xu \& Zhu(2011)]{}    Liang, N., Xu, L., \& Zhu, Z.-H. 2011, A\&A, 527, A11 %(arXiv:1009.6059)
\bibitem [Liang \& Zhu(2010)]{}       Liang, N., \& Zhu, Z.-H. 2010, RAA, in press, arXiv:1010.2681
\bibitem [Lima et al.(2008)]{Lima08}  Lima, J. A. S., Silva, F. E., \& Santos, R. C. 2008, Class and Quantum Gravity, 25, 205006
%\bibitem []{} Linder, E. V. 2010, Phys. Rev. D 81, 127301
%\bibitem []{} Narayan, R., Bartelmann, M., preprint, arXiv:9606001
\bibitem []{} Ma, C., \& Zhang, T. J. 2011, ApJ, 730,  74 %[arXiv:1007.3787]

\bibitem [Nolan et al.(2003a)]{Nolan03a} Nolan, L. A., Dunlop, J. S., Jimenez, R., \& Heavens, A. F. 2003, MNRAS, 341, 464
\bibitem [Nolan et al.(2003b)]{Nolan03b} Nolan, P. L., Tompkins, W. F., Grenier, I. A., \& Michelson, P. F. 2003, ApJ, 597, 615
\bibitem []{} Peebles, P. J. E., \& Ratra, B. 1988a, ApJL, 325, 17
\bibitem []{} Peebles, P. J. E., \& Ratra, B. 1988b, PRD, 37, 3406
\bibitem [Perlmutter et al.(1999)]{Perlmutter99} Perlmutter, S., et al. 1999, ApJ, 517, 565
\bibitem []{} Ratra, B., \& Peebles, P. J. E. 1988, PRD, 37, 3406
\bibitem [Riess(1998)]{Riess98} Riess, A. G., et al. 1998, AJ, 116, 1009
\bibitem [Riess et al.(2004)]{Riess04} Riess, A. G., et al. [Supernova Search Team Collaboration], 2004, ApJ, 607, 665
\bibitem [Riess et al.(2007)]{Riess07} Riess, A. G., et al. 2007, ApJ, 659, 98
\bibitem [Riess et al.(2009)]{Riess09} Riess, A. G., et al. 2009, ApJS, 183, 109
\bibitem [Sahani et al.(2000)]{Sahani00} Sahani, T. D., et al. 2000, PRL, 85, 1162
\bibitem [Samushia \& Ratra(2006)]{Samushia06} Samushia, L., \& Ratra, B. 2006, ApJ, 650, L5
\bibitem [Schaefer(2007)]{Schaefer07} Schaefer, B.~E. 2007, ApJ, 660, 16
\bibitem [Sen \& Scherrer(2008)] {Sen08} Sen, A. A., \& Scherrer, R. J. 2008, PLB, 659, 457
\bibitem [] {} Sereno, M. 2002, A\&A, 393, 757
\bibitem [Simon et al.(2005)]{Simon05} Simon, J., Verde, L., \& Jimenez, R. 2005, PRD, 71, 123001
\bibitem [Stern et al.(2010)]{hz2} Stern, D., et al. 2010, JCAP, 02, 008
\bibitem []{} Tammann, G. A., Sandage, A., \& Reindl, B. 2008, A\&A Rev, 15, 289
\bibitem [Treu et al.(2001)]{Treu01} Treu, T., et al. 2001, MNRAS, 326, 221
\bibitem [Treu et al.(2002)]{Treu02} Treu, T., et al. 2002, ApJL, 564, L13
\bibitem []{} Wan, H. Y., Yi, Z. L., Zhang, T. J., \& Zhou, J, 2007, PLB, 651, 352
\bibitem []{} Wang, S., Li, X. D., \& Li, M. 2010, arXiv:1009.5837
\bibitem []{} Wang, T. S. \& Liang, N. 2010, ScChG, 53, 1720 %[arXiv:0910.5835]
\bibitem [Wei \& Zhang(2007)] {Wei07} Wei, H., \& Zhang, S. N. 2007, PLB, 644, 7
\bibitem []{} Wei, H. 2010, JCAP, 08, 020 [arXiv:1004.4951]
\bibitem []{} Wilson, K. M., Chen, G., \& Ratra, B. 2006, MPLA, 21, 2197
\bibitem [Wood-Vasey et al.(2007)]{Wood07} Wood-Vasey, W. M., et al. 2007, ApJ, 666, 694
\bibitem [Wu \& Yu(2007a)]{Wu07a} Wu, P. X., \& Yu, H. W. 2007a, PLB, 644, 16
\bibitem [Wu \& Yu(2007b)]{Wu07b} Wu, P. X., \& Yu, H. W. 2007b, JCAP, 0703, 015
\bibitem [Wu \& Yu(2010)] {Wu10}  Wu, P. X., \& Yu, H. W. 2010, PLB, 693, 415 %[arXiv:1006.0674]
\bibitem [Xu et al.(2007)]{Xu07} Xu, L. X., Zhang, C. W., Chang, B. R., \& Liu, H. Y. 2008, MPLA, 23, 1939 %[arXiv:0701519]
\bibitem []{} Xu, L. X. \& Wang, Y. T. 2010a, PRD,  82, 043503 %[arXiv:1006.4889]
\bibitem []{} Xu, L. X. \& Wang, Y. T. 2010b, JCAP, 11, 014    %[arXiv:1007.4734]
\bibitem []{} Xu, L. X. \& Wang, Y. T. 2010c, arXiv:1009.0963
\bibitem [Yi \& Zhang(2007)]{Yi07} Yi, Z. L., \& Zhang, T. J. 2007, MPLA, 22, 41
\bibitem []{} Zhai, Z. X., Wan, H. Y., \& Zhang, T. J. 2010, PLB, 689, 8 %[arXiv:1004.2599]
\bibitem []{} Zhang, T. J., \& Ma, C.  2010, arXiv:1010.1307
\bibitem [Zhang \& Zhu(2008)]{Zhang08} Zhang, H. S., \& Zhu, Z.-H. 2008, JCAP, 0803, 007
\bibitem [Zimdahl et al.(2001)]{Zimdahl01} Zimdahl, W., Schwarz, D. J., Balakin, A. B., \& Pavon, D. 2001, PRD, 64, 3501
\bibitem []{} Zhu, Z.-H. 1998, A\&A, 338, 777
\bibitem []{} Zhu, Z.-H., \& Fujimoto, M.-K. 2002, ApJ, 581, 1
%\bibitem []{} Zhu, Z. H. 2004, A\&A, 423, 421
\bibitem []{} Zhu, Z.-H., \& Alcaniz, J. S. 2005, ApJ, 620, 7



\end{thebibliography}
\end{document}